\documentclass[aps,preprint,superscriptaddress]{revtex4-1}

\usepackage[english]{babel}
\usepackage{graphicx}
\usepackage{amsmath}
\usepackage{color}
\usepackage{bm}
\usepackage{xcolor}
\usepackage{hyperref}
\hypersetup{
    colorlinks,
    linkcolor={red!50!black},
    citecolor={blue!50!black},
    urlcolor={blue!80!black}
}

\def\beq{\begin{equation}}
\def\eeq{\end{equation}}

\def\bdm{\begin{displaymath}}
\def\edm{\end{displaymath}}

\def\bea{\begin{eqnarray}}
\def\eea{\end{eqnarray}}

\begin{document}

\title{The van der Waals interaction as the starting point for an
  effective field theory}
\author{Daniel Odell}
\email{dodell@ohio.edu}
\affiliation{Department of Physics and Astronomy, University of Tennessee, Knoxville, TN 37996, USA}
\affiliation{Department of Physics and Astronomy, Ohio University, Athens, OH 45701, USA}
\author{Arnoldas Deltuva}
\affiliation{Institute of Theoretical Physics and Astronomy,
Vilnius University, Saul\.etekio al. 3, LT-10257 Vilnius, Lithuania}
\author{Lucas Platter}
\email{lplatter@utk.edu}
\affiliation{Department of Physics and Astronomy, University of Tennessee, Knoxville, TN 37996, USA}
\affiliation{Physics Division, Oak Ridge National Laboratory, Oak Ridge, TN 37831, USA}
\date{\today}

\begin{abstract}
  We consider the system of three ${}^4$He atoms to assess whether a
  pure van der Waals potential can be used as a starting point for an
  effective field theory to describe three-body processes in ultracold
  atomic systems. Using a long-range van der Waals interaction in
  combination with short-distance two-body counterterms, we analyze
  the dependence of two- and three-body observables on the
  short-distance regulator that is required due to the singular nature of
  the van der Waals interaction. We benchmark our approach with
  results obtained with the {\it realistic} ${}^4$He-${}^4$He LM2M2
  interaction and find good agreement. We furthermore show that in
  this effective field theory approach no three-body force is required
  at leading order and that {\it universal} van der Waals physics
  leads to a universal correlation between three-body observables
  in the absence of an Efimov three-body parameter.
\end{abstract}

\smallskip
\maketitle

\newpage
\section{Introduction}
\label{sec:introduction}
Effective field theories (EFT) have lead to significant advances in the
fields of nuclear and particle physics~\cite{Hammer:2019poc,
  Manohar:2018aog}. In atomic physics, EFTs have
been successfully applied to systems of ultracold atoms with
 large scattering length that also display the so-called
Efimov effect \cite{Efimov:1970zz, Braaten:2004rn}. At leading order
(LO) in this large-scattering-length and short-range EFT (SR-EFT), the
resulting scattering equations are the same as the ones arising from
zero-range interactions \cite{STM57}. Therefore, a three-body
parameter and thereby one experimental three-body datum is required to
make predictions within this framework~\cite{Efimov:1970zz,
  Bedaque:1998km}. Generally, EFTs are low-energy expansions that exploit
a separation of scales (e.g. between large scattering length and range
of the interaction) as an expansion parameter. This implies that every
EFT calculation has an intrinsic uncertainty arising from the
truncation of the expansion which is highly useful for the comparison
with experimental data.

Recently, it was observed that the Efimov three-body parameter in
atomic systems can be derived from the coefficient of the van der
Waals tail of the atom-atom interaction. A number of theoretical works
used various models with long-range van der Waals tails to study these
observations and found that the two-body van der Waals interaction
alone can indeed predict the three-body observables of ultracold gases
with a large scattering length accurately (see Ref. \cite{Naidon_2017}
for a recent review, a discussion of the origin of the so-called van
der Waals universality and a more complete list of references).
However, a recent study~\cite{chapurin2019} has also demonstrated that
there are some caveats to universality that are related to the details
of the atom-atom interaction. These findings immediately raise the
questions whether the van der Waals potential can be the starting
point for atoms whose interaction contains the van der Waals tail and
also what would be the expected uncertainty of a leading order calculation
within such a framework. Van der Waals universality implies
in particular the existence of a short-distance length scale smaller
than $\beta_6$, the length scale associated with the van der Waals
interaction. The size of this short-distance length scale is not
immediately clear, however, it could be determined once certain
aspects of the low-energy expansion have been established.

We note that some effort has already been made to understand (i) the
properties of a two-body system interacting solely through a van der
Waals interaction \cite{PhysRevA.58.1728} and (ii) the contributions
that need to be included beyond leading order
\cite{PhysRevD.78.013001,Brambilla:2017ffe}. However, in this
manuscript, we will focus on the ${}^4$He three-body system that also
displays a large two-body scattering length and try to answer whether
a simple van der Waals potential can be the starting point for an
effective theory description of atomic three-body systems. To answer
this question, one must confront three related questions: Does the
proposed leading order of this new EFT lead to meaningful results that
one can hope to systematically improve upon? What is the 
low-energy EFT expansion parameter and thereby the uncertainty
of a leading-order calculation?  And finally, what are the
required physical parameters that will enter the higher order
calculations?

As a starting point, we will consider the system of three ${}^4$He
atoms. The ${}^4$He interaction leads to a large scattering length and
various potential models have been constructed to reproduce these
features. In the three-body system there are two three-body bound states: a
shallow one, often considered to be an Efimov state associated
with the large scattering length, and a deep state that is considerably
impacted by effective-range corrections. These three-body observables
were calculated many times with the afore-mentioned potentials, and
in particular the so-called LM2M2 model was used in several
theoretical studies\cite{PhysRevA.70.052711, Motovilov:1999iz,
  doi:10.1063/1.481404, Roudnev_2011}. We will therefore use it as a
template for our microscopic underlying theory which serves as the
basis of our EFT approach. This means that selected
 LM2M2 results for observables will be taken to constrain our 
van der Waals EFT model while others are expected to be reproduced by the
 van der Waals EFT with an uncertainty related to its inherent expansion
parameter.

Finally, we should note that this system has also received renewed
attention because a detection of the excited (so-called Efimov)
three-body state was achieved \cite{Kunitski_2015}.

This manuscript is arranged as follows.  In
Sec.~\ref{sec:vdw-interaction}, we present the relevant potential, the
analytical predictions by Gao in the two-body sector, the specifics of
our implementation, and some relevant details of the LM2M2 potential.
In Sec,~\ref{sec:4he-two-body}, we will discuss the details of our
renormalization scheme and the low-energy results obtained in the
two-body sector.  Section~\ref{sec:4he-three-body} contains our
results in the three-body sector and insights into correlations
between the scattering and bound-state regions.  
The results are obtained by solving the corresponding equations in the
 momentum-space partial-wave representation; the  details 
can be found in Ref.~\cite{Odell:2019wjq} and therefore are not discussed here.
Finally, we summarize
our work, draw conclusions, and discuss further advancements in
Sec.~\ref{sec:summary}.

\section{The van der Waals interaction}\label{sec:vdw-interaction}
\subsection{Previous Work}
\label{sec:gao}
In this work, we will consider an interaction that at long distances
has an attractive van der Waals tail of the form
\begin{equation}
  \label{eq:v_vdW}
  V(r) = -\frac{C_6}{r^6}~.
\end{equation}
The van der Waals {\it strength} $C_6$ can be converted into a {\it
  characteristic} length scale $\beta_6\equiv {(mC_6)}^{1/4}$, where
$m$ is the mass of the interacting particles. Gao derived solutions
to the attractive $1/r^6$ potential in
Ref.~\cite{PhysRevA.58.1728}. The bound state wave function of the
state with energy $E$ in partial wave $l$ is written as a linear
combination of two solutions, $f_{E l}(r) $ and $g_{E l}(r)$ of the
van der Waals interaction
\begin{equation}
  \label{eq:GaoSol}
  u_{E l}(r) = A_{E l}\left[f_{E l}(r) - K_l g_{E l}(r)\right]~,
\end{equation}
where $A_{E l}$ is a normalization coefficient and $K_l$ denotes the
so-called short-range $K$-matrix that fixes here an additional boundary
condition on the wave function that is required due to the
potentials singularity at the origin. The precise forms of the
functions $f_{E l}(r)$ and $g_{E l}(r)$ are given in
Ref.~\cite{PhysRevA.58.1728}. For bound states, their asymptotic
form is given by
\begin{align}
  \label{eq:GaoAsym}
  f_{El}(r) \rightarrow {(2\pi \kappa)}^{-1/2} (W_{f-}e^{\kappa r} +
  W_{f+}e^{-\kappa r})~, \nonumber \\
  g_{El}(r) \rightarrow {(2\pi \kappa)}^{-1/2} (W_{g-}e^{\kappa r} +
  W_{g+}e^{-\kappa r})~,
\end{align}
where $\kappa$ represents the bound state momentum and the
coefficients $W_{f\pm,g\pm}$ depend on the energy $E$
and the angular momentum $l$ of the of bound state \cite{PhysRevA.58.1728}.

Requiring Eq.~\eqref{eq:GaoSol} to give normalizable solution implies
that the terms proportional to $e^{\kappa r}$ in
Eq.~\eqref{eq:GaoAsym} cancel and leads to
\begin{equation}
  \label{eq:chi}
  K_l(E) = \chi_l(\Delta) = W_{f-}/W_{g-}~,
\end{equation}
where $\Delta = 2\mu E \beta_6^2/(16\hbar^2)$.

\begin{figure}
  \includegraphics[width=0.9\linewidth]{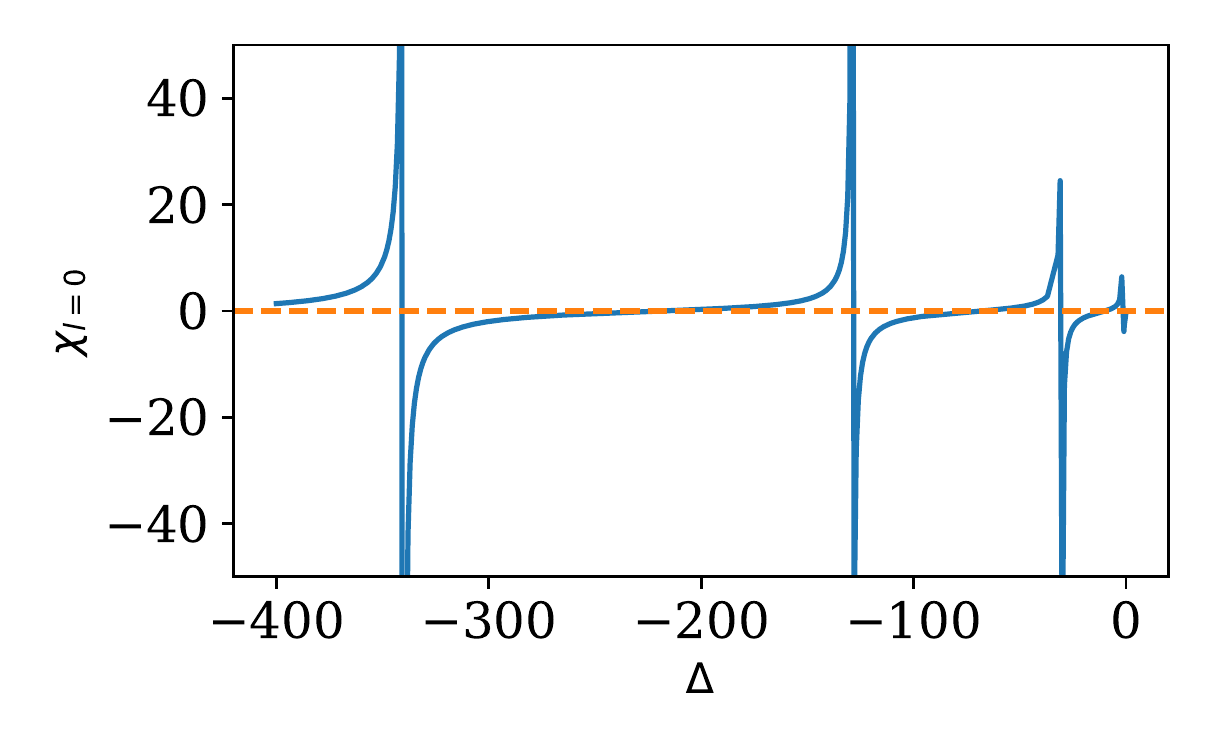}
  \caption{$\chi_{l=0}$ as a function of the dimensionless parameter $\Delta$. The
  solid, blue line is $\chi_{l=0}$. The dashed, orange line is the value of
$\chi_{l=0}$ at the ${}^4\rm{He}_2$ binding energy.}\label{fig:chi}
\end{figure}

The solid line in Fig.~\ref{fig:chi} shows the function
$\chi_{l=0}(\Delta)$ for the $\beta_6$ value associated with the
${}^4$He--${}^4$He interaction. The
intersections between the dashed line and the solid line give the
two-body binding energies in terms of the rescaled energy variable
$\Delta$ once the boundary condition is chosen either by adjusting the
energy ($\Delta$) - position of one the intersections or by adjusting
a scattering observable.

Expressions for the asymptotic solutions at positive energies can be
used to derive expressions for the two-body $t$-matrix and thereby for
the effective range parameters. Gao obtains for the S-wave scattering
length and effective range~\cite{PhysRevA.58.4222}
\begin{equation}
  \begin{split}
    \label{eq:gao_a_and_r}
      a_s & = \frac{2\pi}{{[\Gamma(1/4)]}^2}\frac{K_0(0) -
        1}{K_0(0)}\, \beta_6~, \\
      r_s & \approx \frac{{[\Gamma(1/4)]}^2}{3\pi} \frac{K_0{(0)}^2 +
        1}{{[K_0(0)-1]}^2}\, \beta_6~,
  \end{split}
\end{equation}
where the $K_0(0)$ is evaluated at zero energy (threshold).  The
relation for $r_0$ is truncated under the assumption that the
derivative of the short-range $K$-matrix is small.  In effect, we can
calculate the boundary condition, $K_0(0)$, from $a_s$, and then
calculate $r_s$.

The scattering length is then dependent on the van der Waals length scale,
$\beta_6$, and the short-range $K$-matrix, $K_{l}$, evaluated in the $S$-wave
channel at zero energy.

\subsection{Numerical implementation}
\label{sec:numer-impl}
As a LO approximation of the ${}^4\rm{He}$ system, we take the $C_6$
coefficient from the LM2M2 potential \cite{doi:10.1063/1.460139}, and
account for the short-distance behavior with a single, two-body,
momentum-space counterterm described below. We regulate (cut off) the local part
of the potential at distances well below $R$ with a regulator function $\rho(r;R)$
\begin{equation}
  \label{eq:rho}
  \rho(r;R) = {\left[1 - e^{-{(10r/R)}^2}\right]}^8~,
\end{equation}
such that the full coordinate-space potential is
\begin{equation}
  \label{eq:coordinate_potential}
  V(r)\equiv \rho(r;R) V_6(r)~.
\end{equation}
The local regulator is effective at a distance $R/10$ that is considerably shorter 
compared to the nonlocal regulators described below.
This ensures that cutoff effects are isolated to a single scale --- that there
are no interferences between the local and nonlocal regulators in the
momentum-space potential.
It is also worthwhile noting that other regulators can be used but that their
specific form can influence the rate of convergence with respect to the number
of grid points in numerical calculations.

We will solve for two- and three-body observables in momentum
space. We therefore calculate the momentum space interaction as a regulated
 Fourier transform of the regulated coordinate space version of the
van der Waals interaction
\begin{equation}
  \label{eq:momV}
  \tilde{V}_{l,l^\prime}(p,p') = \tilde{\rho}(p;R) \tilde{\rho}(p';R)~ \frac{2}{\pi} \int_0^\infty dr\,r^2 j_l(pr)
  V(r) j_{l'}(p'r)~,
\end{equation}
where
\begin{equation}
  \label{eq:nonlocal_regulator}
  \tilde{\rho}(p;R) = e^{-(pR/2)^8}~,
\end{equation}
is the nonlocal regulator and $j_l(pr)$ are the spherical Bessel functions.

Once regulated at a short distance, $R$, physical observables acquire
dependence on the arbitrary choice of $R$ which is removed by the
introduction of the counterterm
\begin{equation}
  \label{eq:mom_space_xterm}
  \tilde{\chi}_{l,l'}(p,p';R) = g_l(R)~ p^l (p')^{l'} \tilde{\rho}(p;R) \tilde{\rho}(p';R) ~\delta_{l,l'}~.
\end{equation}
For every value of $R$ the counterterm $g_l(R)$ is readjusted such that the chosen
two-body observable is reproduced. We will refer to the functional
dependence of $g_l(R)$ as renormalization group (RG) flow.
A more detailed discussion of the renormalization scheme can be found in
Subsection~\ref{sec:renormalization}.
Similarly, a more detailed discussion of the calculation of two- and three-body
observables can be found in \cite{Odell:2019wjq}.

\subsection{Details of the LM2M2 Potential}\label{sec:lm2m2-details}

The LM2M2 potential is one of several potentials developed for the
interaction of ${}^4$He atoms~\cite{doi:10.1063/1.460139}. We will use
it here because of a large number of few-body calculations that have
been carried out with this interaction. This potential is a sum of
$r^{-6}$, $r^{-8}$, and $r^{-10}$ terms, each having a separate
strength coefficient. We can quantity how {\it pure} the van der Waals
tail is in the LM2M2 interaction by converting these coefficients
$C_6$, $C_8$ and $C_{10}$ into corresponding length scales. In
Tbl.~\ref{tab:lm2m2betas}, we show  the length scales associated
with each inverse-power-law contribution to the LM2M2 potential. We
can see that the van der Waals tail has the largest length scale. We
expect therefore the features of the trimer states to be dominated by
the van der Waals interaction.

A conservative estimate of the LO theory's uncertainties can be formulated from
the energy scales given in Tbl.~\ref{tab:lm2m2betas}.
The ground-state trimer with 126.4 mK binding energy is the observable farthest from threshold that we
calculate.
The ratio of the binding energy to the energy scale associated with $\beta_8$,
the lowest breakdown scale of the ${}^4\rm{He}-{}^4\rm{He}$ interaction, is
approximately $1/9$.
Therefore the ratio of momentum scales is approximately 1/3.
We use this ratio to conservatively estimate our LO theory uncertainty at 30\%.

In addition, there are three other terms in the LM2M2 potential.
The first is a multiplicative regulator function of the form
\begin{equation}
  \label{eq:lm2m2_reg}
  F(x) = 
  \begin{cases}
    e^{-(D/x-1)^2} & x \leq D \\
    1  & x \geq D
  \end{cases}
\end{equation}
where $x=r/r_m$ and $r_m=2.9695$ \AA.
For the LM2M2, $D=4.2$ \AA.
However, due to the functional form of $F(x)$, the value of $F(x)$ does not drop
below 0.5 until $r<2.5$ \AA.
Therefore, we conclude that the range of the regulator function is comparable to
$\beta_{10}$ which is significantly smaller that the scale of interest,
$\beta_6$.

The second remaining term in the LM2M2 potential is the short-distance
repulsion, which is effectively a Gaussian centered at approximately -8.3~\AA\,
with a 1-$\sigma$ width of 1.5 \AA.
Of course, in the relevant region, $r>0$, this Gaussian function overcomes the
divergences of the inverse-power-law terms.
A conservative estimate for the range of the short-distance repulsion is the
minimum of the total potential, which occurs below 3 \AA, well below $\beta_6$.

The final term to be addressed is a sine function defined in Eq. (A3)
in Ref.~\cite{doi:10.1063/1.460139}.  Its peak value is small enough
relative to the other terms in the potential, therefore its contribution is
insignificant.  To relieve any residual concerns, we point out that
the peak of this term occurs at 3.65 \AA, which is comparable to
$\beta_8$, but again, we emphasize that the magnitude at this peak is
relatively small.

\begin{table}[t] 
	\begin{tabular*}{0.6\textwidth}{@{\extracolsep{\fill}} l c c}
		$n$ & $\beta_n$ & 
		$(m \beta_n^2)^{-1}$ [mK] \\
          \hline
		6                  & 5.38 & 419.18 \\
		8                 & 3.62 & 923.42 \\
		10            & 3.09 & 1272.3 \\
          \hline
	\end{tabular*}
	\caption{The length scales $\beta_6$, $\beta_8$, and
          $\beta_{10}$ and the corresponding energy scales arising
          from the different contributions to the LM2M2 potential}
	\label{tab:lm2m2betas}
\end{table}
\section{The ${}^4$He two-body system}
\label{sec:4he-two-body}

\subsection{Renormalization Scheme}\label{sec:renormalization}

%

We tune the counterterm in Eq. \eqref{eq:mom_space_xterm} in each two-body
partial wave to reproduce arbitrarily chosen low-energy observable(s).
In the $S$-wave channel, $g_{0}(R)$ is tuned to yield a shallow
two-body bound state with the binding energy $B_2=1.31$~mK. 
In the $D$-wave channel $g_2(R)$ is tuned to reproduce
the two-body phase shift at $E=67\rm{mK}~(\Delta=0.01)$
found using the LM2M2 potential.  Finally, the $G$-wave counterterm $g_4(R)$ is
tuned to fit the LM2M2 phase shifts below 15K.  The upper energy limit
of the $G$-wave phase shifts was chosen to avoid low-lying resonances.
The $l=4$ centrifugal barrier of the local van der Waals interaction
constructed here peaks between 14 and 15~K.  Forcing our system to
reproduce the LM2M2 $G$-wave phase shifts over this large region
successfully prevents  resonances, otherwise rapidly moving with $R$,
 from interfering with low-energy three-body observables.  

\begin{figure}[t]
  \begin{center}
    \includegraphics[width=1\textwidth,clip=true]{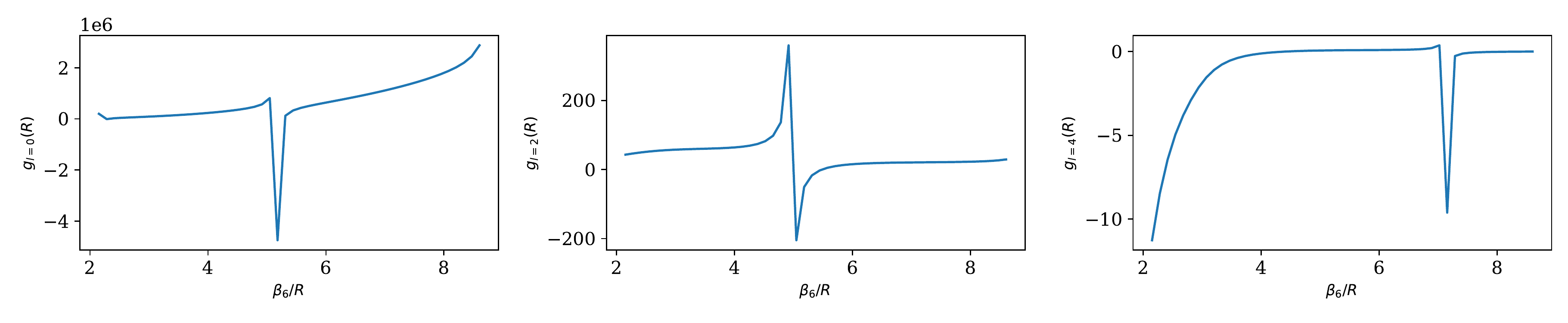}
    \caption{The strengths of the counterterms $g_0(R)$, $g_2(R)$, and $g_4(R)$ as 
    functions of the dimensionless parameter
  $\beta_6/R$. The units of $g_l(R)$ are K$\cdot$\AA${}^{2l+3}$.}\label{fig:rg_flows}
  \end{center}
\end{figure}

This procedure generates a RG {\it flow}, a function that gives the counterterm
coupling strength's dependence on the short-distance regulator scale, $R$, that
is shown for $S$-, $D$-, and $G$-waves in Fig.~\ref{fig:rg_flows}.
 
\subsection{Two-Body Results}\label{sec:two-body-results}

\begin{figure}[t]
  \begin{center}
    \includegraphics[width=0.48\textwidth,clip=true]{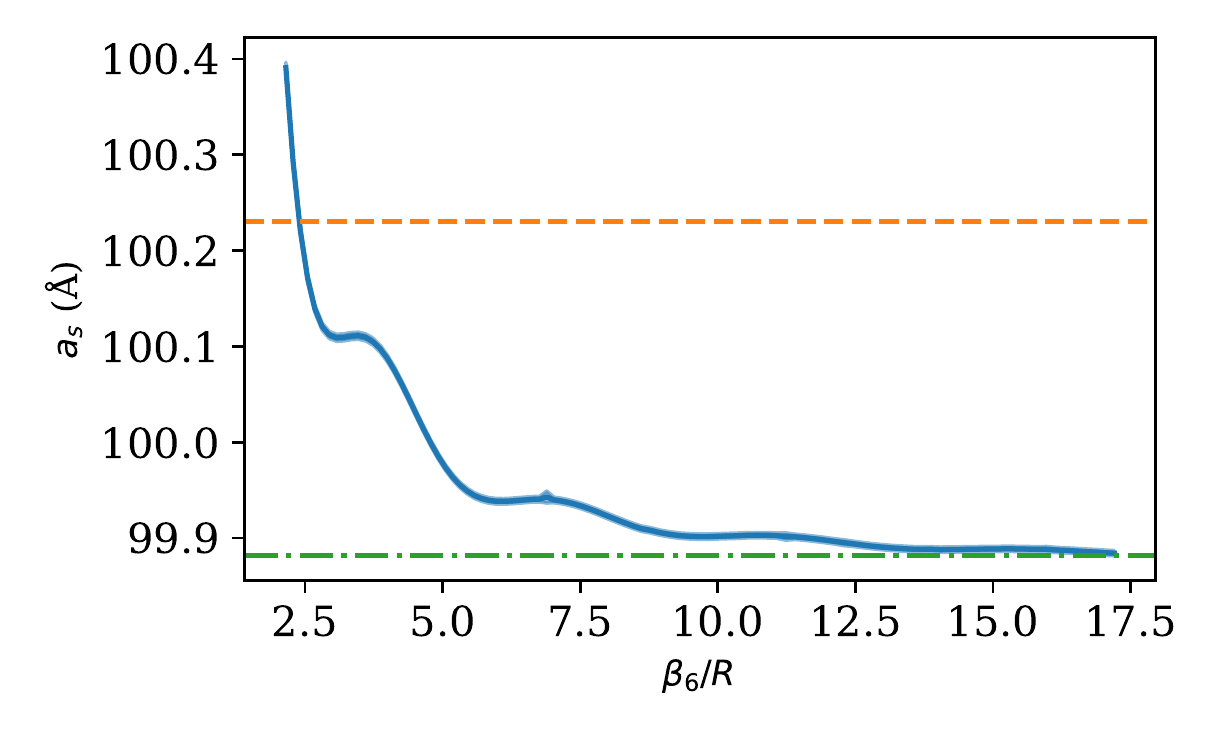} 
    \includegraphics[width=0.48\textwidth,clip=true]{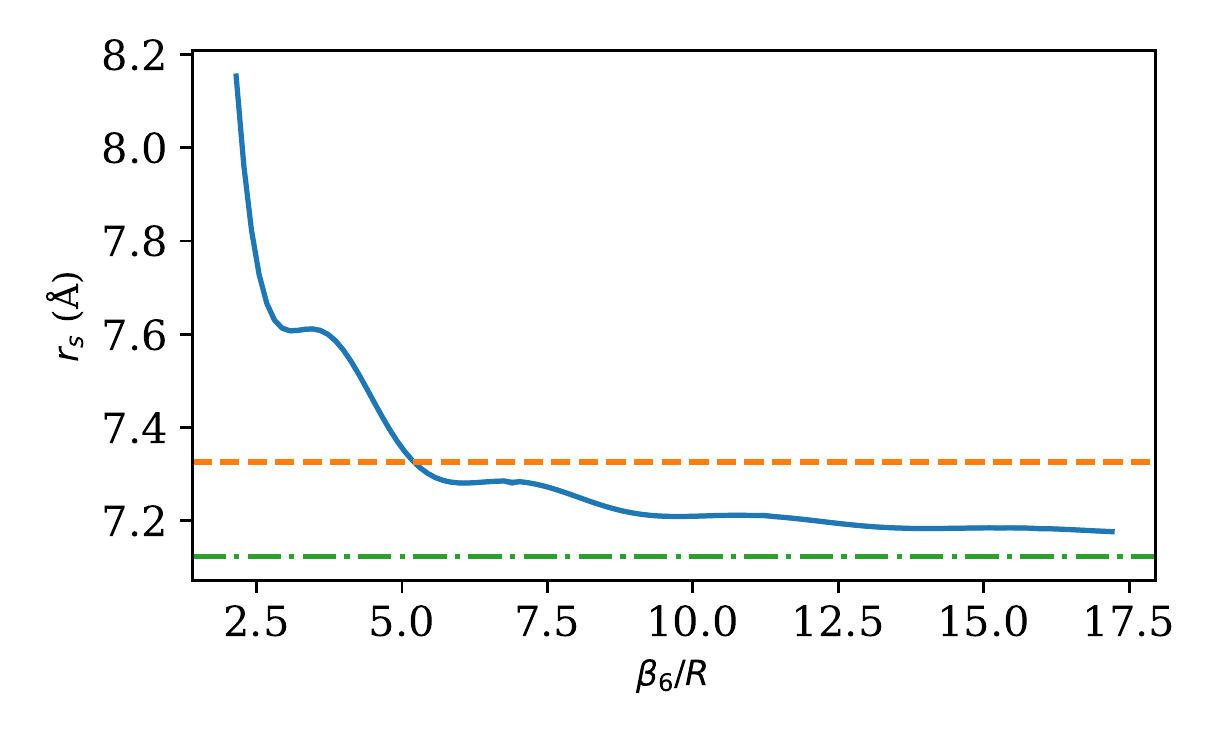} 
  \end{center}
  \caption{The $^{4}$He two-body scattering length (left panel)
    and the effective range (right panel) as  functions of
    $\beta_6/R$. The solid blue lines display the results obtained with
    the renormalized van der Waals potential, the shaded blue band gives the
    numerical uncertainty for $a_s$. The orange dashed lines represent the respective
    results obtained with the LM2M2 potential, while the green dot-dashed lines show Gao's
    prediction based on Eq.~\ref{eq:gao_a_and_r}.}\label{fig:a0}
\end{figure}

To demonstrate the effectiveness of our LO potential to capture the
relevant physical behavior of the ${}^4\rm{He}-{}^4\rm{He}$
interaction, we show in Fig.~\ref{fig:a0} the cutoff dependence of the
S-wave scattering length $a_s$ (left panel) and effective range $r_s$ (right
panel), respectively.
The error bands are generated by the covariance estimates from a non-linear,
least-squares fit.
The LM2M2 results for these parameters are
$a_s = 100.2$~\AA\, and $r_s = 7.33$~\AA\, and are included in both plots
to establish the degree to which the ${}^4\rm{He}-{}^4\rm{He}$
interaction at low energies is characterized by the van der Waals
tail.

In an EFT, we expect the convergence of observables to follow $\mathcal{O}(1/R) =
\mathcal{O}_0\left[1 + \sum_{n=1}^{\infty} c_n (q R)^n\right]$.
As shown in \cite{Odell:2019wjq}, this simple expansion is not always observed
in practice.
Oscillatory functions and non-integer powers of $qR$ can obscure the analytical
description of cutoff dependence --- even at small values of $R$.
In \cite{Odell:2019wjq}, the analysis was conducted for the attractive $1/r^3$
potential.
Here, while similar features can be seen in the figures below, a stronger
singularity makes the calculations much more difficult numerically, rendering the
analysis of the logarithmic derivative inconclusive.
Therefore, asymptotic estimates provided in the following results are based on
qualitative analyses of the short-distance cutoff dependencies.

Figure~\ref{fig:a0} shows the convergence of $a_s$ and $r_s$ with respect to
the dimensionless variable $\beta_6/R$.
The quantities are fit according to the modified effective range expansion derived by
Gao~\cite{PhysRevA.58.4222},
\begin{equation}
  \label{eq:ere}
  k\cot\delta_0 = -\frac{1}{a_s} + \frac{r_s}{2} k^2 + c_3 k^3 + c_4 k^4 +
  \mathcal{O}(k^4\ln{k})~.
\end{equation}
Estimated values for $a_s$ and $r_s$ in the $\beta_6/R\rightarrow\infty$ limit
are given in Table~\ref{tab:asymptotic_values}.
In comparison to the LM2M2 results, $a_s$ is within 0.5\% and $r_s$ within 2\%.
The relative difference between $a_s$ and the Gao prediction is on the order of
10${}^{-5}$.
The effective range differs by less than a percent.
The scattering length is given {\it for free} in that universality, of the
generic kind, guarantees this result.
The agreement of $r_s$ with the LM2M2 and Gao's prediction~\cite{PhysRevA.58.4222}
is the result of van der Waals universality.

The agreement between our results and Gao's predictions is noticeably better for
$a_s$ than for $r_s$.
We point out a few caveats associated with this observation.
First, $r_s$ can be sensitive to the energy range included in the fit to
Eq.~\eqref{eq:ere}, an arbitrary choice.
Similarly, that fit is also sensitive to the number of terms treated in
Eq.~\eqref{eq:ere}.
Finally, the analytical prediction in \cite{PhysRevA.58.1728} 
for $r_s$ includes a term
proportional to the energy derivative of the $K$-matrix at zero energy.
This term is assumed to be small and therefore not included in the evalution
here.



\begin{figure}[t]
  \begin{center}
    \includegraphics[width=0.8\textwidth,clip=true]{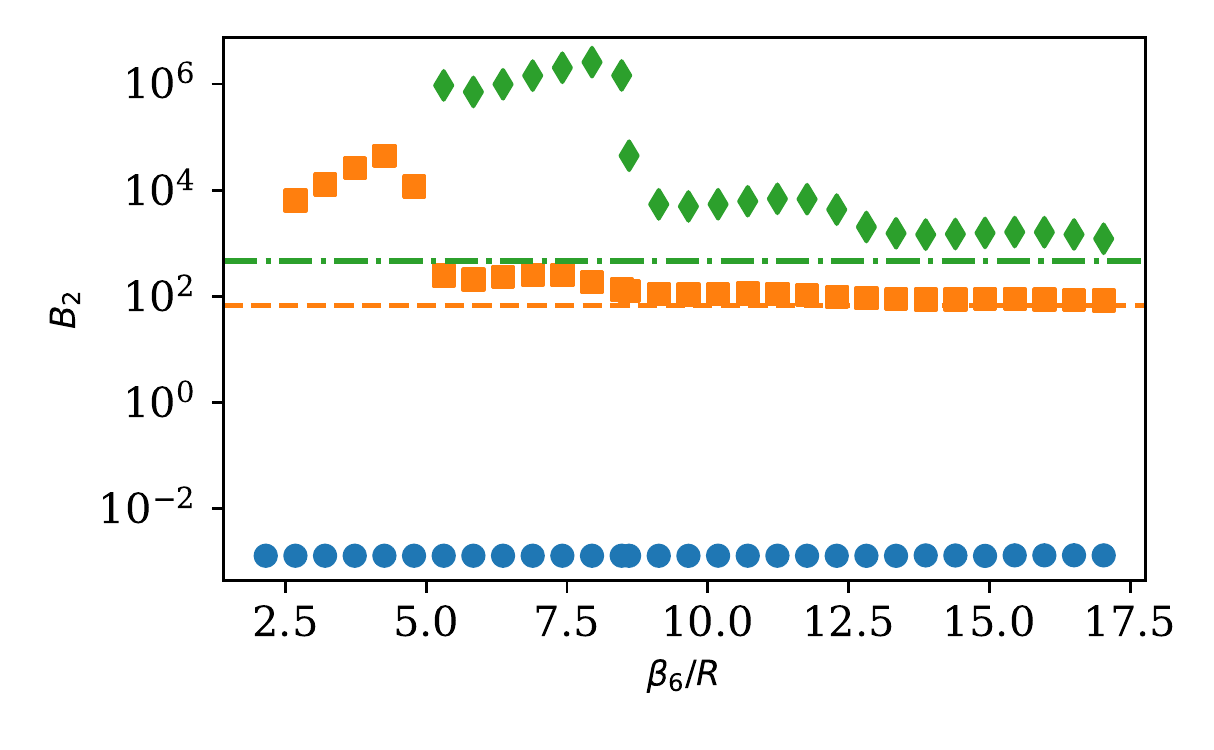} 
  \end{center}
  \caption{Evolutions of the two-body s-wave spectrum with a decreasing short-distance cutoff $R$.
    The shallowest state (blue circles) is fixed and shown together
    with the next two states in the spectrum. The first one (orange squares) is
    comparied to the Gao prediction for the next deeper state (dashed, orange
    lines). The second one (green diamonds) is similarly compared to the Gao
    prediction (green, dash-dotted line). Predictions are discussed in
    \ref{sec:gao}.}\label{fig:swave_spec}
\end{figure}

As a further demonstration of the agreement between our results and the Gao
predictions, we show the two-body spectrum in Fig.~\ref{fig:swave_spec}.
As $\beta_6/R$ increases, the degree to which the van der Waals tail dominates
the long-range behavior increases, moving our results towards those predicted by
Gao.
Indeed, the binding energies of deeper two-body states are approaching  the
Gao's predictions --- precisely the behavior we expect.

\begin{figure}[t]
  \begin{center}
    \includegraphics[width=0.48\textwidth,clip=true]{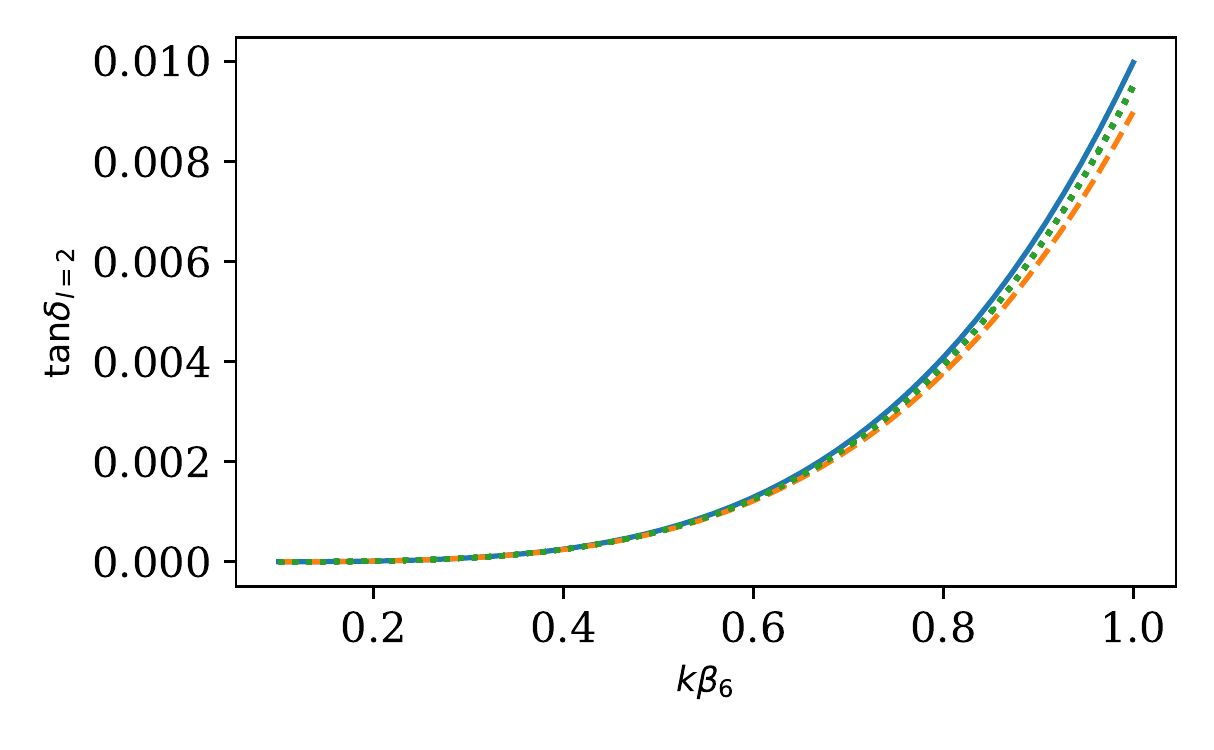} 
    \includegraphics[width=0.48\textwidth,clip=true]{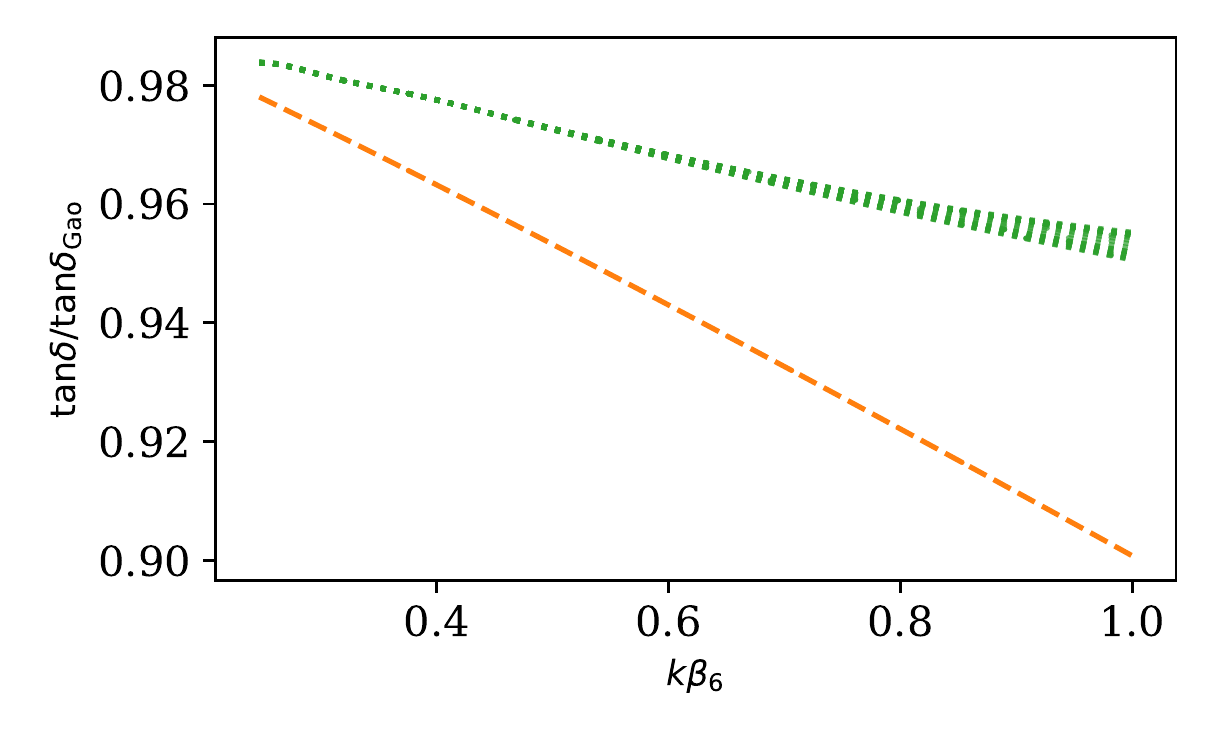} 
  \end{center}
  \caption{[Left panel] The van der Waals (green, dotted lines) and LM2M2
    (orange, dashed line) $d$-wave phase shifts, given in terms of $\tan\delta$,
    are compared to Gao's predictions (blue, solid line). The different
    overlapping van der Waals lines correspond to different values of $R$.
    [Right panel] The ratio of $\tan\delta_{l=2}$ to the prediction by Gao as
    a function of $k\beta_6$. Where applicable, the colors and linestyles are as
    in the left panel.}\label{fig:phase_shifts_dwave} \end{figure}

Furthermore, Gao predicts that for higher partial waves,
\begin{equation}
  \label{eq:gao_tandelta_l_ge_2}
  \tan\delta_{l\ge 2} = \frac{3\pi}{32(l+1/2)\left[(l+1/2)^2-4\right]
  \left[(l+1/2)^2-1\right]} (k\beta_6)^4~.
\end{equation}
In the left panel of Fig.~\ref{fig:phase_shifts_dwave} we compare the
$D$-wave phase shifts of our pure van der Waals interaction with the result
obtained using the LM2M2 potential and the Gao
prediction~\cite{PhysRevA.58.4222}.
The different lines shown
correspond to different values of the short-distance regulator
$R$. The results agree very well at low momenta while they start to
deviate at larger momenta. However, we note that the scale on which the
phaseshift is shown is relatively small. A relative comparison is
shown in the right panel of Fig.~\ref{fig:phase_shifts_dwave} where
the differences between small values of $\tan\delta$ are highlighted.
The van der Waals system, as expected, consistently stays closer to the
predicted behavior whereas the LM2M2 results quickly deviate
presumably due to significant contributions from terms in the
potential dominating at the short distances being probed at higher
energies.

\section{The ${}^4$He three-body system}
\label{sec:4he-three-body}

We will now calculate three-body observables using the van der Waals
interaction tuned to reproduce ${}^4$He two-body parameters.  We
emphasize that the theory developed at this order is uninformed by
${}^4\rm{He}_3$ observables which is a very different starting point
in comparison to the SR-EFT.

\subsection{Binding energies}
\begin{figure}[t]
  \begin{center}
    \includegraphics[width=0.48\textwidth,clip=true]{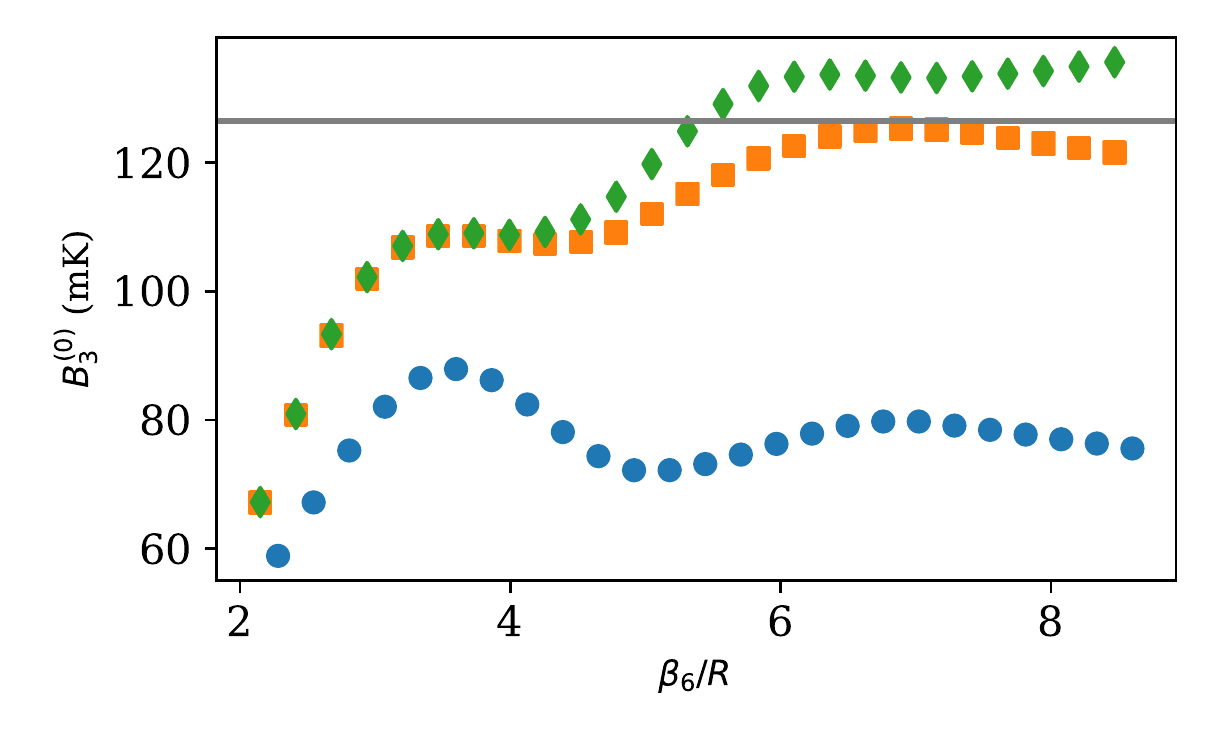}
    \includegraphics[width=0.48\textwidth,clip=true]{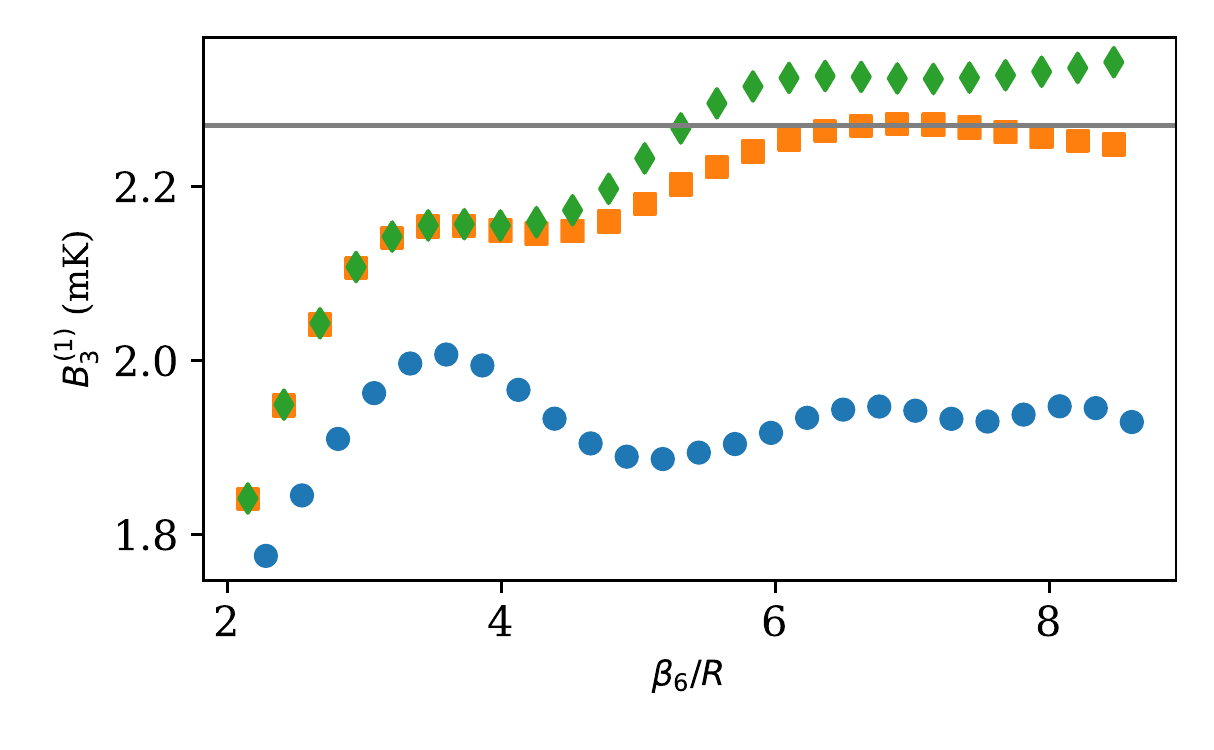} 
\end{center}
\caption{Three-body binding energies for the ground (left) and excited (right)
  states as functions of $\beta_6/R$. The blue circles are the calculated
  energies using $S$-wave only. Orange squares include  $D$-wave in addition, and green
  diamonds include also $G$-wave.}
\label{fig:binding0}
\end{figure}

We first calculate the ground state binding energy of the ${}^4$He
trimer. The LM2M2 potential prediction for this state is 126.4~mK
\cite{Roudnev_2011}. In Fig.~\ref{fig:binding0} we show the binding
energy of the trimer ground state as a function of the dimensionless
parameter $\beta_6/R$. The blue circles give the results when only
$S$-wave two-body interaction is included in the calculation, orange squares
 include also the $D$-wave, and green diamonds include $G$-wave as well.
Since $\beta_6$ the is the characteristic length scale of the van der Waals interaction, we
expect that only values of $\beta_6/R > 1$ capture the important features of the
van der Waals interaction.
As indicated by the figure, in the $\beta_6/R\rightarrow\infty$ and
$l_{\max}\rightarrow\infty$ limits, a converged value near (within 10\%) the
LM2M2 result is obtained.

The LM2M2 potential leads to a binding energy of 2.27~mK for the
excited trimer state~\cite{Roudnev_2011}.  From an energy-scale point
of view, the excited state appears to be a better observable to study
for the purposes of EFTs: For the SR-EFT, it is a state whose
binding energy is close to the energy scale of the two-body bound
state and that is thereby clearly within the range of convergence of this
EFT. For our van der Waals EFT it is well separated from the energy scales
associated with $\beta_8$ and $\beta_{10}$  shown in
Table~\ref{tab:lm2m2betas}. The regulator dependence of the excited
state is shown in the right panel of Fig.~\ref{fig:binding0}.  The
convergence toward the LM2M2 result is similar to that of the
ground-state trimer and further demonstrates the accuracy of
describing the ${}^4\rm{He}-{}^4\rm{He}$ interaction with a van der
Waals EFT.

Estimates for the three-body binding energies in the
$\beta_6/R\rightarrow\infty$ limit are given in
Table~\ref{tab:asymptotic_values}.  Contributions from partial waves
where $l>4$ are neglected.  Based on the rapid decrease in the
contribution to $B_3^{(n)}$ going from $l_{\max}=2$ to $l_{\max}=4$,
we expect higher partial waves to have even less of an impact.  This
effect has already been observed in~\cite{Deltuva_2015} for the LM2M2
potential.

\subsection{Atom-dimer scattering}
\begin{figure}[t]
  \begin{center}
    \includegraphics[width=0.8\textwidth,clip=true]{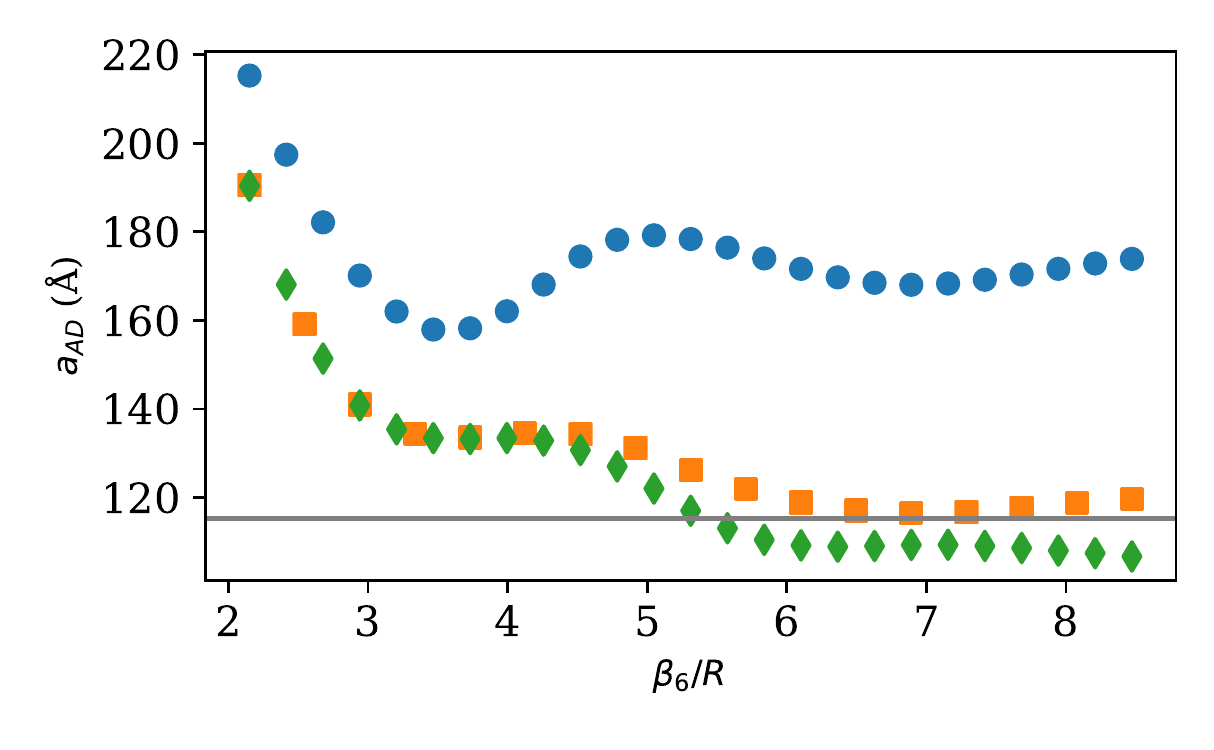}
  \end{center}
  \caption{Atom-dimer scattering length $a_{AD}$ as a function of
    $\beta_6/R$. The horizontal solid line gives the result of
    Ref.~\cite{Roudnev_2011}. 
    Symbols and colors are as  in Figure~\ref{fig:binding0}}\label{fig:a21}
\end{figure}

The atom-dimer scattering properties of the $^{4}$He have been
considered previously with various realistic atom-atom
interactions.
In Ref.~\cite{Roudnev_2011}, a value of 115.22~\AA~is obtained for the LM2M2
potential. In Fig.~\ref{fig:a21}, we show the convergence of the atom-dimer
scattering length with respect to $\beta_6/R$. Symbols and colors are as they
appear in Fig.~\ref{fig:binding0}. 
We observe again that at least $S$- and $D$-waves have to be included to obtain
good agreement with the results of the LM2M2 potential.
Assuming again that the regulator dependence scales as $\beta_6/R$, we
determine the atom-dimer scattering length to be $105 \pm 10$~\AA\,
in the limit of $\beta_6/R \rightarrow \infty$.
Though not shown in the figure, we obtained stable numerical results for $S$-
and $D$-wave calculations of $a_{AD}$ for values of $\beta_6/R\lesssim17$ and
verified that the convergence is stable.

We have also calculated the atom-dimer effective range and find $r_{AD} = 85 \pm
10$~\AA\, but note that the extraction of this observable is difficult due to
its strong regulator dependence.
This result has to be compared to the LM2M2 result of 79.0~\AA\, given in
Ref.~\cite{Lazauskas_2006}.
The convergence of $B_3^{(0)}$, $B_3^{(1)}$, and $a_{AD}$ with respect to the
$\beta_6/R\rightarrow\infty$ limit indicates that a three-body force is not
required at LO to obtain renormalized results.

\begin{table}[t]
\begin{tabular*}{0.6\textwidth}{@{\extracolsep{\fill}}  l c c }
  $\mathcal{O}$ & $\mathcal{O}((\beta_6/R)_{\max})$ & $\delta\mathcal{O}$ \\ 
    \hline 
    $a_s$ (\AA) & 99.9 & 0.1 \\  
    $r_s$ (\AA) & 7.2 & 0.1 \\
    $B_3^{(0)}$ (mK) & 135 & 10 \\
    $B_3^{(1)}$ (mK) & 2.3 & 0.1 \\
    $a_{AD}$ (\AA) & 105 & 10 \\
    $r_{AD}$ (\AA) & 85 & 10 \\
  \hline
    \label{tab:asymptotic_values}
\end{tabular*}
\caption{Estimates for various observables ($\mathcal{O}$) and conservative
estimates of their associated uncertainties ($\delta\mathcal{O}$) based on the
convergence behavior shown in the previous figures.} \end{table} 

\subsection{Universal correlations and a comparison  to the short-range EFT}
\label{sec:uni_corr}
\begin{figure}[t]
  \begin{center}
    \includegraphics[width=0.48\textwidth,clip=true]{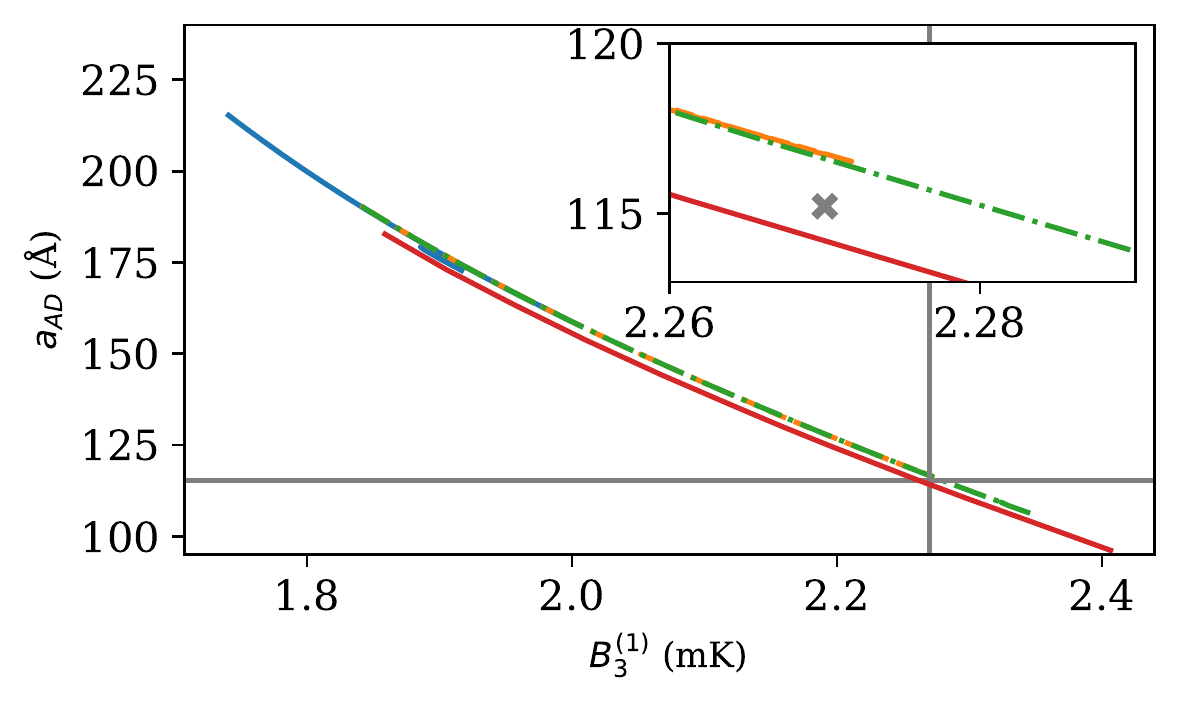}
        \includegraphics[width=0.48\textwidth,clip=true]{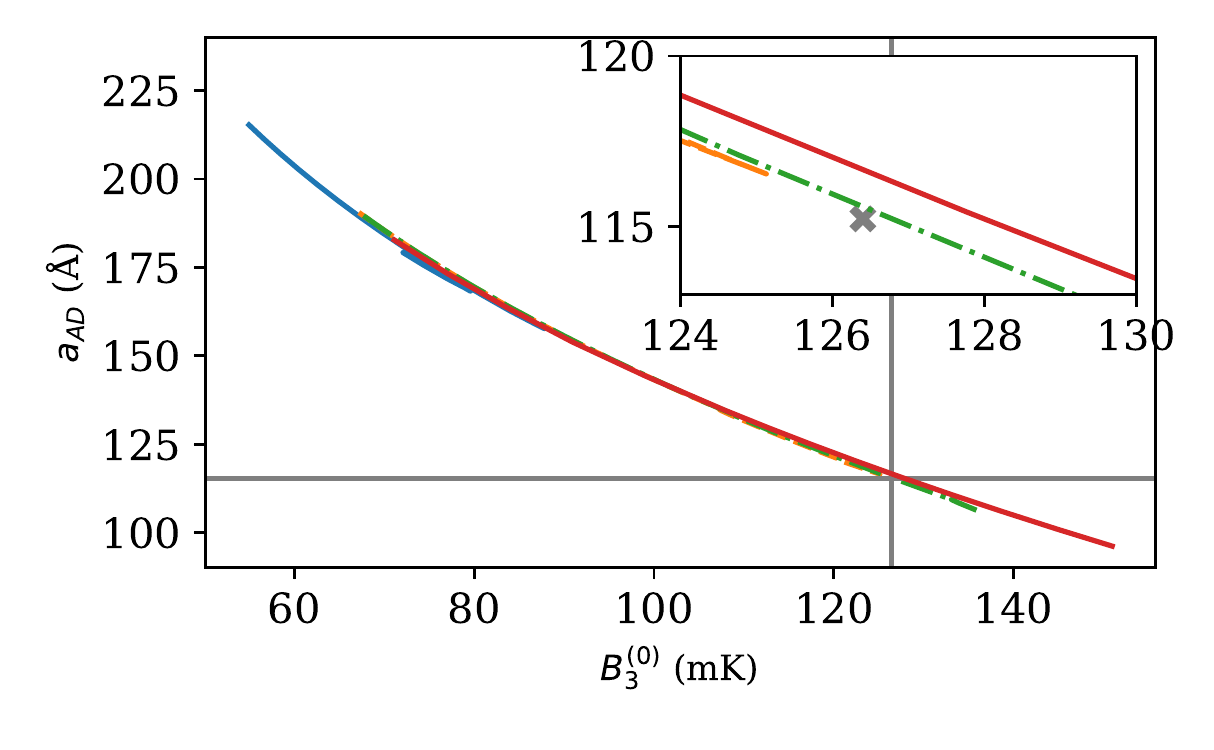} 
  \end{center}
  \caption{Left panel: The correlation between the excited state
    binding energy $B_3^{(1)}$ and the atom-dimer scattering length $a_{AD}$.
    Right panel: The correlation between the trimer ground state energy and the
    atom-dimer scattering length $a_{AD}$. [Both panels] $S$-wave only results
    are shown as a blue, solid line.  Results including $D$- and $G$-waves are
    shown as orange, dashed and green, dot-dashed lines, respectively. The red,
    solid line is the prediction from SR-EFT.}\label{fig:a21_b3_1_corr}
\end{figure}

A correlation between the neutron-deuteron scattering length and the
triton binding energy was observed in calculations with phaseshift
equivalent interactions and is known as the Phillips
line~\cite{Phillips:1968zze}. In the EFT with contact interactions,
this correlation can easily be reproduced by varying the three-body
parameter~\cite{Bedaque:1998kg}. Reference
\cite{Braaten:2002jv} used the SR-EFT to analyze universal aspects of this
system and displayed the Phillips line for the ${}^4$He
system produced at LO. In Refs.~\cite{Platter:2006ev, Ji:2012nj}, SR-EFT
calculations were carried out at next-to-leading order (NLO) and
next-to-next-to-leading order (N2LO) for the $^{}4$He system by
including the effects of a finite effective range and it was found
that the convergence behavior of the EFT expansion is poor for the
ground state.

In Fig.~\ref{fig:a21_b3_1_corr}, we display the correlation between $a_{AD}$ and
both $B_3^{(1)}$ and $B_3^{(0)}$ values obtained by changing the regulator $R$.
The results show an approximately linear correlation. However, more importantly:
results that were obtained with different numbers of included partial wave
channels fall on nearly the same correlation line.  In addition to our own
results, we include the results corresponding to the values reported in
\cite{Roudnev_2011} shown as grey horizontal and vertical lines. The
intersection of the lines representing those results falls in excellent
agreement with our own calculations.
In Fig.~\ref{fig:a21_b3_1_corr}, we show the so-called Phillips line predicted
by SR-EFT as a red, solid line.
The agreement with this Phillips line traced out by our calculations is quite
good.
For the excited state, the differences between the predicted correlations and
the results of \cite{Roudnev_2011} are qualitatively the same.
In the case of the ground state, the correlation from van der Waals EFT is
noticably closer to intersection of the results reported in \cite{Roudnev_2011}.
It is particularly remarkable that the results for calculations that
include different numbers of partial waves in the two-body subsystem lie on the
same line.
The reason for this is that the low-energy scattering features of higher partial
waves are not relevant to obtain the correlations between observables in the
${}^4$He three-body sector.
At distances where the van der Waals interaction dominates the centrifugal
barrier the short-range attraction is the same in all partial waves.
The degree to which this attraction is exposed determines the location on the
${}^4$He Phillips line.

\section{Summary}
\label{sec:summary}
In this work, we considered the three-body system composed of
${}^4\rm{He}$ atoms as a starting point for a description of few-body
observables with an EFT whose leading order is the
long-range van der Waals interaction. We analyzed for the first time the
dependence of three-body observables on the short-distance regulator
employed in calculations with the van der Waals potentials. Our numerical
results converge and become independent of the short-distance
regulator which leads to the conclusion that no three-body force is
required when the van der Waals interaction with a two-body counterterm is
considered to be the LO of an EFT. We provided values with uncertainty
estimates for the trimer binding energies and for the atom-dimer
scattering length in the limit of zero short-distance regulator.

We demonstrated also that this leading order calculation provides a
good description of the ${}^4$He trimer system, {\it i.e.}, the binding
energies of the two trimer bound states and the atom-dimer scattering
length are very close to the results obtained with the LM2M2
potential. In agreement with previous calculations using $^{4}$He
potentials, we observe that higher two-body partial waves have an
important impact (specifically the $l=2$ contribution) on the
three-body observables but that the size of their contribution
decreases rapidly with increasing $l$. Also, comparing the result for
the deep trimer state obtained in SR-EFT and van der Waals EFT indicates that
the latter apparently contains important finite range contributions
as the result is significantly closer to the potential result. This is
encouraging as higher-order calculations with the SR-EFT indicate that the
deep trimer state is outside of the radius of convergence of the
SR-EFT~\cite{Ji:2012nj}. We furthermore found that the inclusion of
higher partial waves does not shift significantly the so-called Phillips line,
the correlation between the atom-dimer scattering length
and three-body binding energy obtained by varying the short-distance
regulator, but rather move along the Phillips line.

Two important developments should be carried out in the near
future. First, the structure and impact of higher order corrections
needs to be analyzed. As a first step, momentum-dependent contact
interactions can be introduced in the S-wave channel. It would then be
necessary to determine the observables to fit this additional
low-energy parameter to, e.g. low-energy phaseshifts in a certain
energy range. Developments related to the use of the so-called
modified effective range expansion in the context of
EFTs~\cite{Steele:1998zc} and recently developed prescriptions for the
construction of ordering schemes for singular
interactions~\cite{Long:2007vp} make this avenue particularly
promising. Second, this approach should be used to analyze the
three-body spectrum for a variable scattering length, {\it i.e.}, for
systems close to a Feshbach resonance. This will provide further
insights on the uncertainties for three-body observables and universal
relations established for the van der Waals universality.

\begin{acknowledgments}
We thank H.-W. Hammer, D.~R.~Phillips, and S. K\"onig for useful discussions.
This work has been supported by the National Science Foundation under Grant
No. PHY-1555030, the Office of Nuclear Physics, U.S. Department of Energy
under Contract No. DE-AC05-00OR22725, and the National Nuclear Security
Administration No.~DE-NA0003883.
This work used the Extreme Science and Engineering Discovery Environment
(XSEDE), which is supported by National Science Foundation grant number
ACI-1548562, allocation number TG-PHY200044~\cite{6866038}.
\end{acknowledgments}

\bibliographystyle{apsrev}

\end{document}